\documentclass[11pt,english]{article}
\pdfoutput=1
\usepackage{jheppub}
\usepackage{amsmath}
\usepackage{amssymb}
\usepackage{amsfonts}

\usepackage{graphicx}

\newcommand{\beq}{\begin{equation}}
\newcommand{\eeq}{\end{equation}}
\newcommand{\ft}[2]{{\textstyle\frac{#1}{#2}}}
\newcommand{\fft}[2]{\frac{#1}{#2}}
\newcommand{\nn}{{\nonumber}}

\begin{document}

\preprint{MCTP-16-07, YITP-SB-16-7}

\title{Holographic Lifshitz fermions and exponentially suppressed spectral weight}

\author[a,b]{Youngshin Kim}

\affiliation[a]{C.N. Yang Institute for Theoretical Physics, Department of Physics and Astronomy,\\
Stony Brook University, Stony Brook, NY 11794--3840, USA}

\author[b]{and James T. Liu}

\affiliation[b]{Michigan Center for Theoretical Physics, Randall Laboratory
of Physics,\\
 The University of Michigan, Ann Arbor, MI 48109--1040, USA}

\emailAdd{youngshin.kim.1@stonybrook.edu}  \emailAdd{jimliu@umich.edu}

\abstract{The absence of fixed momentum excitations in a theory with Lifshitz scale invariance
gives rise to exponential suppression of spectral weight in the low-frequency limit.  In the holographic
dual, this suppression arises as a consequence of a tunneling barrier that decouples the horizon
from the boundary.   We compute the spin-1/2 holographic Green's function and show that the
form of the barrier is identical to that of the scalar case.
We furthermore demonstrate that the suppression factor is universal in the $\hat\omega\to0$ limit where
$\hat\omega=\omega/|\vec k|^z$.  In particular, it depends only on $\hat\omega$ and the critical
exponent $z$, and is independent of scaling dimension and spin.}

\maketitle

%%%%%%%%%%
\section{Introduction}

Following the remarkable successes of AdS/CFT and its applications to strongly coupled
gauge theories, there has been much recent interest in developing a similar program for
non-relativistic duals.  In particular, quantum critical phenomena with the scaling symmetry
\begin{equation}
t\to\lambda^z t,\qquad \vec x\to\lambda\vec x,
\label{eq:lifsca}
\end{equation}
where $z$ is the dynamical exponent, can be described in the framework of Lifshitz
holography with a bulk geometry given by \cite{Kachru:2008yh}
\begin{equation}
ds_{n+2}^2=-\left(\fft{L}r\right)^{2z}dt^2+\left(\fft{L}r\right)^2\left(d\vec x_n^2+dr^2\right).
\label{eq:lifmet}
\end{equation}
Here we have chosen the radial coordinate so that the boundary scaling (\ref{eq:lifsca}) is
associated with the transformation $r\to\lambda r$.

Much of the analysis of Lifshitz holography follows from the well-developed methods of
relativistic AdS/CFT.  However, there are some key differences.  Firstly, while the bulk
metric (\ref{eq:lifmet}) may in some sense be viewed as a generalization of the Poincar\'e
patch of AdS, it in fact has a tidal singularity at the horizon
\cite{Kachru:2008yh,Hartnoll:2009sz,Copsey:2010ya,Horowitz:2011gh}.
There are of course various approaches to handling this mild singularity
\cite{Harrison:2012vy,Bao:2012yt,Bhattacharya:2012zu,Knodel:2013fua}.  However, a more
substantial difference appears
in the causal structure of the Lifshitz spacetime.  In particular, while null geodesics in an
AdS background will reach the boundary in finite coordinate time, only radial null geodesics
will do so in a Lifshitz background \cite{Keeler:2013msa}.  Thus, in the classical limit,
boundary probes carrying transverse momentum decouple from the interior of the Lifshitz
spacetime.

It was demonstrated in \cite{Keeler:2013msa} that this decoupling persists at the quantum
level in that bulk scalar modes carrying large transverse momentum become `trapped' at the
horizon in the sense that they leave only an exponentially small imprint on the boundary.
As a consequence, the smearing function that maps from the boundary to the bulk is
ill-defined in a strict sense, and can at most be viewed only as a distribution.  Moreover,
this feature of Lifshitz holography remains even when the tidal singularity at the horizon
is removed.

The origin of the trapped scalar modes can be seen by investigating the effective
potential for the radial part of the Klein-Gordon equation.  In contrast with a pure AdS
background, the Lifshitz geometry gives rise to a tunneling barrier for sufficiently large
spatial momentum.  This in turn leads to exponential suppression in the radial
wavefunction that decouples the corresponding mode from the boundary.  This suppression
also shows up in the holographic Green's function in the form of an exponentially
suppressed spectral weight in the limit of large transverse momentum
\cite{Faulkner:2010tq,Hartnoll:2012wm,Keeler:2014lia}.

For a scalar operator with Lifshitz scaling dimension $\nu$, a WKB computation demonstrated
that for $\hat\omega\ll\nu^{-(z-1)}$ the exponential suppression of spectral weight is of the form
\cite{Faulkner:2010tq}
\begin{equation}
\chi=2\,\mbox{Im}\, G_R\sim\exp\left[-\fft{\sqrt\pi\Gamma(1/(2z-2))}{z\Gamma(z/(2z-2))}\left(\fft1{\hat\omega}\right)^{1/(z-1)}\right],
\label{eq:chitun}
\end{equation}
where $\hat\omega=\omega/|\vec k|^z$.  On the gravitational side of the duality, this
suppression is a general feature of the Lifshitz geometry, while on the field theory side it is a
direct consequence of Lifshitz scaling.  In particular, exponential suppression in the limit
$\hat\omega\to0$ is simply a statement that excitations do not fall below the dispersion relation
$\omega\sim|\vec k|^z$ \cite{Hartnoll:2012rj}.  (When $z=1$, the suppression is in fact complete,
as $\chi=0$ for $\omega<|\vec k|$.)

While exponential suppression in the low-frequency limit is a robust feature of Lifshitz scaling, the
actual suppression factor may {\it a priori} be operator and spin dependent.  Thus we are motivated
to investigate the spin-1/2 fermion Green's function in a pure Lifshitz geometry.  We show that,
although the square of the Dirac equation in the bulk does not coincide with the scalar Klein-Gordon
equation, the suppression factor remains identical to that of the scalar in (\ref{eq:chitun}).
Fermion correlators in AdS/CFT were
investigated in \cite{Henningson:1998cd,Mueck:1998iz,Henneaux:1998ch}, and further developed in
\cite{Iqbal:2009fd,Liu:2009dm,Faulkner:2009wj}.  Exponential suppression of fermion spectral weight has
previously been noted in
\cite{Faulkner:2010tq,Iizuka:2011hg,Hartnoll:2011dm,Iqbal:2011in,Cubrovic:2011xm,Gursoy:2012ie}, but
is absent for vanishing bulk fermion mass \cite{Alishahiha:2012nm,Fang:2012pw}.  Exponential suppression
in the current-current correlator was demonstrated in \cite{Hartnoll:2012wm} and at non-zero temperature
in the $T\to0$ limit in \cite{Son:2002sd,Horowitz:2009ij,Basu:2009xf,Hartnoll:2012rj}.

In section~\ref{sec:dirac}, we examine the Dirac
equation and the holographic computation of the Green's function in a Lifshitz background.  Then, in
section~\ref{sec:wkb}, we compute the fermion spectral function using the WKB approximation.  Taking
the low frequency limit, $\hat\omega\to0$, then reproduces the exponential suppression indicated in
(\ref{eq:chitun}).  Finally, we conclude in section~\ref{sec:dis} with an argument that the suppression
factor in (\ref{eq:chitun}) is universal, regardless of spin and scaling dimension.

%%%%%%%%%%
\section{The Dirac equation in a Lifshitz background} 
\label{sec:dirac}

We consider a free Dirac particle in the bulk, with equation of motion
\begin{equation}
(\Gamma^M\nabla_M-m)\psi=0.
\label{eq:diraceqn}
\end{equation}
The covariant derivative acting on $\psi$ is given by $\nabla_M=\partial_M
+\fft14\omega_M{}^{\bar N\bar P}\Gamma_{\bar N\bar P}$, where $\omega_M{}^{\bar N\bar P}$
is the spin connection, and bars indicate tangent space indices.  To proceed, we
note that the spin connection for a metric of the form
\begin{equation}
ds^2=-e^{2A(r)}dt^2+e^{2B(r)}d\vec x_n^2+e^{2C(r)}dr^2,
\end{equation}
is especially simple.  This allows us to rewrite the Dirac equation as
\begin{equation}
(\Gamma^M\partial_M-m)\tilde\psi=0,
\label{eq:partialD}
\end{equation}
where $\tilde\psi=e^{\fft12(A+nB)}\psi$.  Note that, while the Dirac operator is written in terms of partial
derivatives, it still corresponds to a curved background since the Dirac matrices are
written in curved space.  In particular
\begin{equation}
\Gamma^t = e^{-A} \Gamma^{\bar{t}},\quad
\Gamma^i = e^{-B} \Gamma^{\bar{i}},\quad \Gamma^r = e^{-C} \Gamma^{\bar{r}},
\end{equation}
where $i$ goes over the spatial coordinates.

Since the boundary theory is translationally invariant, we work in momentum space, and
hence assume a
plane wave solution of the form $\tilde\psi(t,\vec{x}) = e^{i(\vec{k}\cdot \vec{x}-\omega t)} f(r)$.
Inserting this into (\ref{eq:partialD}) and rearranging gives
\begin{equation}
[\Gamma^{\bar{r}} \partial_r -me^C
-i(\omega e^{C-A} \Gamma^{\bar{t}}- k_ie^{C-B}\Gamma^{\bar{i}})]f(r) = 0.
\end{equation}
At this point, it is worth recalling some basic facts about fermions in AdS/CFT.
Specializing to the Lifshitz metric (\ref{eq:lifmet}), where $e^{-A} = (r/L)^z$ and
$e^{-B} = e^{-C} = r/L$, the Dirac equation becomes
\begin{equation}
\left[\Gamma^{\bar{r}} \partial_r -\fft{mL}r
-i\left(\omega\left(\fft{r}L\right)^{z-1} \Gamma^{\bar{t}}- k_i\Gamma^{\bar{i}}\right)\right]f(r) = 0.
\label{eq:diracint}
\end{equation}
Taking $z>1$, the boundary (i.e.\ $r\to0$) behavior is governed by $(\Gamma^{\bar r}\partial_r
-mL/r)f(r)\approx0$, which is solved by
\begin{equation}
f^{(1)}\sim r^{-mL},\qquad f^{(2)}\sim r^{mL},
\label{eq:bdasy}
\end{equation}
where $f^{(1)}=-\Gamma^{\bar r}f^{(1)}$ and $f^{(2)}=\Gamma^{\bar r}f^{(2)}$
have definite $\Gamma^{\bar r}$ eigenvalues.
Taking $m>0$ (which we assume throughout), this demonstrates that $f^{(1)}$ corresponds to
the source, as it is non-normalizable as $r\to0$, while $f^{(2)}$ corresponds to the response.

Unlike in the scalar case, the Dirac equation is first order, and can be thought of as relating
half of the spinor components to the other half.  From a holographic point of view, this implies
that the CFT operator has half the spinor components as that of the bulk Dirac field \cite{Iqbal:2009fd}.
For an even dimensional boundary theory, $f^{(1)}$ couples to a chiral fermionic operator (with
$\Gamma^{\bar r}$ serving as the chirality matrix).  For an odd dimensional boundary theory,
the spinor components are halved simply because the Dirac representation is halved by
going down in dimension.

Motivated by (\ref{eq:bdasy}), we now split $f(r)$ into source ($\eta$) and response ($\xi$)
terms according to
\begin{equation}
f=\eta-i\Gamma^{\bar t}\xi,\qquad
\Gamma^{\bar r}\eta=-\eta,\qquad
\Gamma^{\bar r}\xi=-\xi.
\label{eq:fetaxi}
\end{equation}
The Dirac equation (\ref{eq:diracint}) then reduces to the system of coupled linear equations
\begin{eqnarray}
(\partial_r+mL/r)\eta&=&(\omega(r/L)^{z-1}-k_i\Gamma^{\bar t\bar i})\xi,\nn\\
(\partial_r-mL/r)\xi&=&-(\omega(r/L)^{z-1}+k_i\Gamma^{\bar t\bar i})\eta.
\end{eqnarray}
These are matrix equations for the chiral spinors $\eta$ and $\xi$.  To proceed, we note
that $k_i\Gamma^{\bar t\bar i}$ has eigenvalues $\pm k$.  This allows us make the further
projections
\begin{equation}
\eta_\pm = P_\pm\eta,\qquad\xi_\pm=P_\pm\xi,
\end{equation}
where
\begin{equation}
P_\pm=\fft12\left(1\pm\fft{k_i\Gamma^{\bar t\bar i}}k\right),\qquad k\equiv|\vec k\,|.
\end{equation}
It is important to keep in mind that this is separate from (and commutes with) the
$\Gamma^{\bar r}$ projection used above.  The end result is a pair of coupled equations
(one for $\eta_+$ and $\xi_+$ and another for $\eta_-$ and $\xi_-$)
\begin{eqnarray}
(\partial_r+m/r)\eta_\pm&=&(\omega r^{z-1}\mp k)\xi_\pm,\nn\\
(\partial_r-m/r)\xi_\pm&=&-(\omega r^{z-1}\pm k)\eta_\pm,
\label{eq:1order}
\end{eqnarray}
where we have set $L=1$.
At this stage, we may note that component equations (\ref{eq:1order}) are invariant under the interchange
$\omega\to-\omega$ along with $\eta_\pm\to\eta_\mp$ and $\xi_\pm\to-\xi_\mp$.  Since this relates positive
and negative $\omega$, we may focus on $\omega>0$ and use this symmetry to handle the case of
negative $\omega$.

The above equations can be recast in second-order form
\begin{eqnarray}
\left(\partial_r^2-\fft{\nu_+^2-1/4}{r^2}+\omega^2r^{2(z-1)}-k^2\right)\eta_\pm
-\fft{(z-1)\omega r^{z-2}}{\omega r^{z-1}\mp k}\left(\partial_r+\fft{m}r\right)\eta_\pm&=&0,\nn\\
\left(\partial_r^2-\fft{\nu_-^2-1/4}{r^2}+\omega^2r^{2(z-1)}-k^2\right)\xi_\pm
-\fft{(z-1)\omega r^{z-2}}{\omega r^{z-1}\pm k}\left(\partial_r-\fft{m}r\right)\xi_\pm&=&0,
\label{eq:2order}
\end{eqnarray}
where $\nu_\pm=m\pm1/2$.  Finally, we transform to Schr\"odinger form by defining
\begin{equation}
\eta_\pm=\sqrt{\omega r^{z-1} \mp k}\, \Psi^{(1)}_\pm, \quad
\xi_\pm = \sqrt{\omega r^{z-1} \pm k}\, \Psi^{(2)}_\pm.
\label{eq:trans}
\end{equation}
We then obtain
\begin{equation}
-\partial_r^2\Psi^{(1)}_\pm + V^{(1)}_\pm\Psi^{(1)}_\pm = 0, \quad
-\partial_r^2\Psi^{(2)}_\pm + V^{(2)}_\pm\Psi^{(2)}_\pm = 0,
\label{eq:Sch}
\end{equation}
with effective potentials \cite{Gursoy:2012ie}
\begin{eqnarray}
V^{(1)}_\pm &=& \frac{\nu_+^2-1/4}{r^2} - \omega^2 r^{2(z-1)} + k^2
+(z-1)\omega r^{z-3}\left(\fft{\nu_+}{\omega r^{z-1} \mp k}
+\fft{z-1}4\fft{\omega r^{z-1} \pm2k}{(\omega r^{z-1} \mp k)^2}\right) ,\nn\\
V^{(2)}_\pm& = & \frac{\nu_-^2-1/4}{r^2} - \omega^2 r^{2(z-1)} + k^2
+(z-1)\omega r^{z-3}\left(-\fft{\nu_-}{\omega r^{z-1} \pm k}
+\fft{z-1}4\fft{\omega r^{z-1} \mp2k}{(\omega r^{z-1} \pm k)^2}\right).\nn\\
\label{eq:v}
\end{eqnarray}

Since these Schr\"odinger potentials were essentially obtained from the square of the
Dirac equation, it is interesting to compare them to the corresponding potential for a
bulk scalar.  We see that the terms in (\ref{eq:v}) not proportional to $z-1$ are universal,
and match those of the scalar potential
\begin{equation}
V_0=\fft{\nu_0^2-1/4}{r^2}-\omega^2r^{2(z-1)}+k^2,
\label{eq:Vscalar}
\end{equation}
while the remaining terms are specific to the Dirac equation.  Provided $z>1$, the
boundary ($r\to0$) behavior is governed by the first terms in (\ref{eq:v}), which provides
the proper scaling dimension of the dual operator.  On the other side, the horizon
behavior is dominated by $-\omega^2 r^{2(z-1)}$.  The additional terms in the potential
will contribute to the bulk wavefunction, but will not substantially modify its
qualitative features.

There is, however, a subtlety in the approach of using second order equations, and that is
that the denominators in (\ref{eq:v}) or even (\ref{eq:2order}) may blow up somewhere in
the bulk.  Since we focus on $\omega>0$, this happens in the $\Psi^{(1)}_+$ and
$\Psi^{(2)}_-$ equations.  (The magnitude of the wavevector $\vec k$ is always
non-negative).  Note also that the square-root factors in
(\ref{eq:trans}) can become imaginary on one side of the singularity.

Of course, we expect that the solution to the first order equations (\ref{eq:1order}) ought to
be well behaved at $\omega r^{z-1}-k=0$.  Moreover, for $\omega>0$, the second order
equations for $\Psi^{(1)}_-$ and $\Psi^{(2)}_+$ are well-behaved for $r>0$.  This provides
us with a well-defined procedure for solving the system.  We first solve the effective
Schr\"odinger problem for $\eta_-$ and $\xi_+$, and then use the first order
equations (\ref{eq:1order}) to obtain $\eta_+$ and $\xi_-$.  This also has the benefit of
guaranteeing that the first order equations are consistently solved without a doubling
of the degrees of freedom that would arise from independent second order equations.

\subsection{Extracting the holographic Green's function}
\label{sec:extrac}

We follow the procedure of \cite{Iqbal:2009fd} to obtain the fermionic Green's function from the
solution to the bulk Dirac equation.  The starting point is the boundary asymptotics of the Dirac field
$\psi=r^{\fft12(z+n)}f(r)$, which may be obtained by dropping the $\omega$ term in (\ref{eq:diracint}).
As highlighted by (\ref{eq:bdasy}), we split $\psi$ into components $\psi^{(1)}$ and $\psi^{(2)}$ with definite
$\Gamma^{\bar r}$ eigenvalues.  The Dirac equation can be converted to second-order form, with
asymptotic behavior
\begin{eqnarray}
\psi^{(1)}&=&Ar^{\fft12(z+n)-m}+Br^{\fft12(z+n)+m+1},\nn\\
\psi^{(2)}&=&Cr^{\fft12(z+n)-m+1}+Dr^{\fft12(z+n)+m},
\label{eq:asybeh}
\end{eqnarray}
as $r\to0$.  Here $A$, $B$, $C$ and $D$ are constant spinors that are related by the first order
Dirac equation
\begin{equation}
C=\fft{ik_i\Gamma^{\bar i}}{2m-1}A,\qquad B=\fft{ik_i\Gamma^{\bar i}}{2m+1}D.
\end{equation}
The Green's function $G$ then relates the normalizable mode ($D$) to the non-normalizable one ($A$)
\begin{equation}
D=G(i\Gamma^{\bar t})A.
\label{eq:Gdef}
\end{equation}
As usual, the retarded Green's function $G_R$ is obtained by taking infalling boundary conditions at
the horizon.

Now suppose we have a solution to the effective Schr\"odinger problem (\ref{eq:Sch}) for $\Psi_-^{(1)}$
and $\Psi_+^{(2)}$ of the form
\begin{eqnarray}
\Psi_-^{(1)}&=&\alpha_{(1)}r^{-m}+\beta_{(1)}r^{m+1},\nn\\
\Psi_+^{(2)}&=&\alpha_{(2)}r^{-m+1}+\beta_{(2)}r^m,
\label{eq:albe}
\end{eqnarray}
as $r\to0$.  We may then use (\ref{eq:trans}) and (\ref{eq:1order}) to translate this back to expressions for
$\eta$ and $\xi$
\begin{eqnarray}
\eta_+&=&\fft{2m-1}k\alpha_{(2)}\sqrt{k}r^{-m}-\fft{k}{2m+1}\beta_{(2)}\sqrt{k}r^{m+1},\nn\\
\eta_-&=&\alpha_{(1)}\sqrt{k}r^{-m}+\beta_{(1)}\sqrt{k}r^{m+1},\nn\\
\xi_+&=&\alpha_{(2)}\sqrt{k}r^{-m+1}+\beta_{(2)}\sqrt{k}r^m,\nn\\
\xi_-&=&-\fft{k}{2m-1}\alpha_{(1)}\sqrt{k}r^{-m+1}+\fft{2m+1}k\beta_{(1)}\sqrt{k}r^{m}.
\end{eqnarray}
Using the relation (\ref{eq:fetaxi}) then allows us to extract the coefficients
\begin{eqnarray}
D&=&-i\Gamma^{\bar t}\ft12(1-\Gamma^{\bar r})\sqrt{k}\left(P_+\beta_{(2)}+\fft{2m+1}kP_-\beta_{(1)}\right),\nn\\
A&=&\ft12(1-\Gamma^{\bar r})\sqrt{k}\left(\fft{2m-1}kP_+\alpha_{(2)}+P_-\alpha_{(1)}\right),
\end{eqnarray}
where we have made the projections explicit.  Comparing this with (\ref{eq:Gdef}) then gives
\begin{equation}
G_R=\ft12(1+\Gamma^{\bar r})\left(P_+G_{(1)}+P_-G_{(2)}\right),
\label{eq:Gfromab}
\end{equation}
where
\begin{equation}
G_{(1)}=-\fft{2m+1}k\fft{\beta_{(1)}}{\alpha_{(1)}},\qquad
G_{(2)}=-\fft{k}{2m-1}\fft{\beta_{(2)}}{\alpha_{(2)}}.
\label{eq:G12}
\end{equation}
Note that, while $\alpha_{(i)}$ and $\beta_{(i)}$ are spinors, since there is a single effective Schr\"odinger
equation governing each pair $\{\alpha_{(1)},\beta_{(1)}\}$ and $\{\alpha_{(2)},\eta_{(2)}\}$, the
ratios $\beta_{(i)}/\alpha_{(i)}$ are well defined.  Furthermore, the overall $\Gamma^{\bar r}$ projection is
related to the fact that the boundary spinor degrees of freedom are half that of the bulk.

To summarize, the procedure for obtaining the holographic Green's function for $\omega>0$
is first to solve the effective
Schr\"odinger equation (\ref{eq:Sch}) for $\Psi_-^{(1)}$ and $\Psi_+^{(2)}$, and then to extract the
quantities $\alpha_{(i)}$ and $\beta_{(i)}$ from the boundary asymptotics according to (\ref{eq:albe}).
The Green's function then has the block diagonal form (\ref{eq:Gfromab}).  For $\omega<0$, we use
the symmetry properties of (\ref{eq:1order}) to map
\begin{equation}
G_{(1)}(-\omega,k)=-G_{(2)}^*(\omega,k),\qquad
G_{(2)}(-\omega,k)=-G_{(1)}^*(\omega,k).
\label{eq:Gnegw}
\end{equation}
%

%%%%%%%%
\section{The holographic Green's function in the WKB approximation}
\label{sec:wkb}

We are now ready to examine the holographic Green's function.  As we have seen above, we
first solve the Dirac equation in the bulk, which can be transformed into the equivalent
Schr\"odinger-like equations (\ref{eq:Sch}) for $\Psi_-^{(1)}$ and $\Psi_+^{(2)}$. The effective
potentials $V_-^{(1)}$ and $V_+^{(2)}$ given in (\ref{eq:v}) can be rewritten as
\begin{eqnarray}
V_-^{(1)}&=&\fft{\nu_+^2-1/4}{r^2}-\omega^2r^{2\delta}+k^2+\fft{\delta}{r^2}(\omega r^\delta)
\left(\fft{\nu_+}{\omega r^\delta+k}+\fft{\delta}4\fft{\omega r^\delta-2k}{(\omega r^\delta+k)^2}
\right),\nn\\
V_+^{(2)}&=&\fft{\nu_-^2-1/4}{r^2}-\omega^2r^{2\delta}+k^2+\fft{\delta}{r^2}(\omega r^\delta)
\left(-\fft{\nu_-}{\omega r^\delta+k}+\fft{\delta}4\fft{\omega r^\delta-2k}{(\omega r^\delta+k)^2}
\right),
\label{eq:Spotd}
\end{eqnarray}
where $\delta\equiv z-1>0$.  The first three terms in the potential are common to the scalar case, while the
terms in the parentheses arise from squaring the Dirac equation.  Note that the potential is not analytic
in $r$ for non-integer $\delta$.

The behavior of the potentials near the boundary is given by
\begin{eqnarray}
V_-^{(1)}(r\to0)\approx\fft{\nu_+^2-1/4}{r^2}+\fft{\omega\delta(\nu_+-\delta/2)/k}{r^{2-\delta}}
+\cdots.\nn\\
V_+^{(2)}(r\to0)\approx\fft{\nu_-^2-1/4}{r^2}-\fft{\omega\delta(\nu_-+\delta/2)/k}{r^{2-\delta}}+\cdots.
\end{eqnarray}
These potentials remain dominated by the $1/r^2$ term, which leads to the expected
power-law boundary asymptotics $\Psi_-^{(1)}\sim r^{1/2\pm\nu_+}$ and
$\Psi_+^{(2)}\sim r^{1/2\pm\nu_-}$ as indicated in (\ref{eq:albe}).  The horizon behavior,
on the other hand, is universal, and arises from the divergent term
\begin{equation}
V(r\to\infty)\sim -\omega^2 r^{2\delta}.
\end{equation}
The horizon behavior is then of the form
\begin{equation}
\Psi(r\to\infty)\sim r^{-\delta/2}e^{\pm i\omega r^z/z}.
\end{equation}
The retarded Green's function is obtained by taking the positive sign, corresponding to
infalling boundary conditions.

In general, it is not possible to obtain a closed-form solution to the Schr\"odinger equation
with potential given by (\ref{eq:Spotd}).  However, we can search for tunneling barriers,
and hence regions of exponential suppression, using the WKB approximation.  Following
Ref.~\cite{Keeler:2014lia}, which analyzed the scalar case, we assume the potential $V$
admits a single turning point $r_0$ with $V(r_0)=0$.  The first order WKB wavefunction is then given by
\begin{equation}
\Psi_{\rm WKB}(r)=\begin{cases}\displaystyle
\fft1{\hat V(r)^{\fft14}}\left[F\exp\left(\int_r^{r_0}dr'\sqrt{\hat V(r')}\right)+G\exp\left(-\int_r^{r_0}dr'\sqrt{\hat V(r')}\right)\right],&r<r_0;\\
\displaystyle
\fft1{|\hat V(r)|^{\fft14}}\left[a\exp\left(i\int_{r_0}^rdr'\sqrt{|\hat V(r')|}\right)+b\exp\left(-i\int_{r_0}^rdr'\sqrt{|\hat V(r')|}\right)\right],&r>r_0,
\end{cases}
\label{eq:wkb}
\end{equation}
where $\hat V(r)=V(r)+1/4r^2$.  This shift arises when applying the WKB
method to a potential that blows up as $1/r^2$, and corresponds to making the shift
$\nu_\pm^2-1/4 \to\nu_\pm^2$ in the first term of the potential in (\ref{eq:Spotd}).
See appendix~\ref{app:WKB} for details.

The boundary behavior, encoded by the constants $F$ and $G$, is related to the horizon
behavior, encoded by the constants $a$ and $b$, via the WKB connection formulae
\begin{equation}
\begin{pmatrix}F\\
G\end{pmatrix}=\begin{pmatrix}e^{-i\pi/4} & e^{i\pi/4} \\
\fft12e^{i\pi/4} & \fft12e^{-i\pi/4}\end{pmatrix}\begin{pmatrix}a\\
b\end{pmatrix}.
\label{eq:WKBbc}
\end{equation}
In order to compute the holographic Green's function (\ref{eq:Gfromab}), we need to
relate the boundary coefficients $\alpha_{(i)}$ and $\beta_{(i)}$ defined in (\ref{eq:albe}) to
the WKB coefficients $F$ and $G$.  Using the boundary behavior $\hat V\approx\nu^2/r^2$, we see that
\begin{equation}
\Psi_{\rm WKB}(r\to0)\sim Fc_1r^{1/2-\nu}+Gc_2r^{1/2+\nu},
\end{equation}
where $c_1$ and $c_2$ are constants obtained by integrating up to the classical turning point.
This demonstrates that $F$ is related to the non-normalizable mode with coefficient $\alpha_{(i)}$
and $G$ is related to the normalizable mode with coefficient $\beta_{(i)}$.  This allows us to
obtain
\begin{eqnarray}
\alpha_{(i)}&=&\lim_{r\to0}\fft{Fr^{\nu-1/2}}{\hat V(r)^{\fft14}}\exp\left(\int_r^{r_0}dr'\sqrt{\hat V(r')}\right),\nn\\
\beta_{(i)}&=&\lim_{r\to0}\fft{Gr^{-\nu-1/2}}{\hat V(r)^{\fft14}}\exp\left(-\int_r^{r_0}dr'\sqrt{\hat V(r')}\right).
\label{eq:abWKB}
\end{eqnarray}

At this point, it is worth noting that while the term proportional to $F$ in the WKB approximation
is dominated by the non-normalizable mode, in fact it may contain a sub-dominant (and
in general non-analytic) piece related to the normalizable mode.  As discussed in
\cite{Keeler:2014lia}, the WKB approximation is unreliable in extracting this subdominant
behavior.  However, this only affects the real part of the holographic Green's function,
and in particular the imaginary part remains under control.  In any case, what this means
is that only the component of $\beta_{(i)}$ in (\ref{eq:abWKB}) that is orthogonal to
$\alpha_{(i)}$ in the complex plane is to be trusted.

For the retarded Green's function, we take infalling boundary conditions, which corresponds
to setting $b=0$ in (\ref{eq:WKBbc}).  The holographic Green's function is then given by
(\ref{eq:Gfromab}) where
\begin{equation}
\fft{\beta_{(i)}}{\alpha_{(i)}}=\fft{i}2\lim_{r\to0}r^{-2\nu}\exp\left(-2\int_r^{r_0}dr'\sqrt{\hat V(r')}\right).
\label{eq:boa}
\end{equation}
Here $\nu$ is taken to be $\nu_+$ and $\nu_-$, while $\hat V(r)$ is $\hat V^{(1)}_-$ and $\hat V^{(2)}_+$,
respectively, for $i=1$ and $i=2$.  Note that this quantity is manifestly imaginary, and
contributes to the spectral function.  The real part of the Green's function is actually
non-vanishing, but cannot be obtained by this WKB method.

Since the WKB approximation is reliable for the imaginary part of the Green's function, we focus on
the spectral function, defined as
\begin{equation}
\chi=-\fft1\pi\mbox{Im}\,\mbox{Tr}\, G_R=-\fft{2^{\lfloor n/2\rfloor}}{2\pi}\mbox{Im}(G_{(1)}+G_{(2)}),
\end{equation}
where $G_{(1)}$ and $G_{(2)}$ are given in (\ref{eq:G12}), and may be obtained from (\ref{eq:boa}).
(The normalization factor is related to the bulk dimension $n+2$.)
For $k\ne0$, we may transform the expression (\ref{eq:boa}) into a dimensionless WKB integral by
letting $x=kr$.  The result is then
\begin{eqnarray}
\mbox{Im}\,G_{(1)}&=&-\fft{2m+1}2k^{2m}\lim_{\epsilon\to0}\epsilon^{-2\nu_+}
\exp\left(-2\int_\epsilon^{x_0}dx'\sqrt{\hat V_-^{(1)}(x')}\right),\nn\\
\mbox{Im}\,G_{(2)}&=&-\fft1{2(2m-1)}k^{2m}\lim_{\epsilon\to0}\epsilon^{-2\nu_-}
\exp\left(-2\int_\epsilon^{x_0}dx'\sqrt{\hat V_+^{(2)}(x')}\right),
\label{eq:green-1st}
\end{eqnarray}
where
\begin{eqnarray}
\hat V_-^{(1)}(x)&=&\fft{\nu_+^2}{x^2}-\hat\omega^2 x^{2\delta}+1+\fft\delta{x^2}\hat\omega x^\delta
\left(\fft{\nu_+}{\hat\omega x^\delta+1}+\fft\delta4\fft{\hat\omega x^\delta-2}{(\hat\omega x^\delta+1)^2}
\right),\nn\\
\hat V_+^{(2)}(x)&=&\fft{\nu_-^2}{x^2}-\hat\omega^2 x^{2\delta}+1+\fft\delta{x^2}\hat\omega x^\delta
\left(-\fft{\nu_-}{\hat\omega x^\delta+1}+\fft\delta4\fft{\hat\omega x^\delta-2}{(\hat\omega x^\delta+1)^2}
\right).
\label{eq:xpot}
\end{eqnarray}
Here we have defined the scale-invariant quantity $\hat\omega=\omega/k^z$.

The expressions (\ref{eq:green-1st}) are valid for $\hat\omega>0$.  The behavior for $\hat\omega<0$,
may be obtained using (\ref{eq:Gnegw}).  While in the above we have focused on first order WKB, the
analysis can be extended to higher order as well.  The second order correction is presented in
Appendix~\ref{app:WKB}, and as an example we show a comparison of the first and second order
approximations with the exact numerical solution in Fig.~\ref{fig:g1}.  Details of how we obtained the
numerical solution are given in Appendix~\ref{app:exact}.  Although there is considerable deviation
of the first order WKB result from the exact solution for negative $\hat\omega$, it agrees well in
the exponential suppression region around $\hat\omega\approx0$.

%%%%
\begin{figure}[t]
\centering
\includegraphics[width=1\textwidth,origin=c]{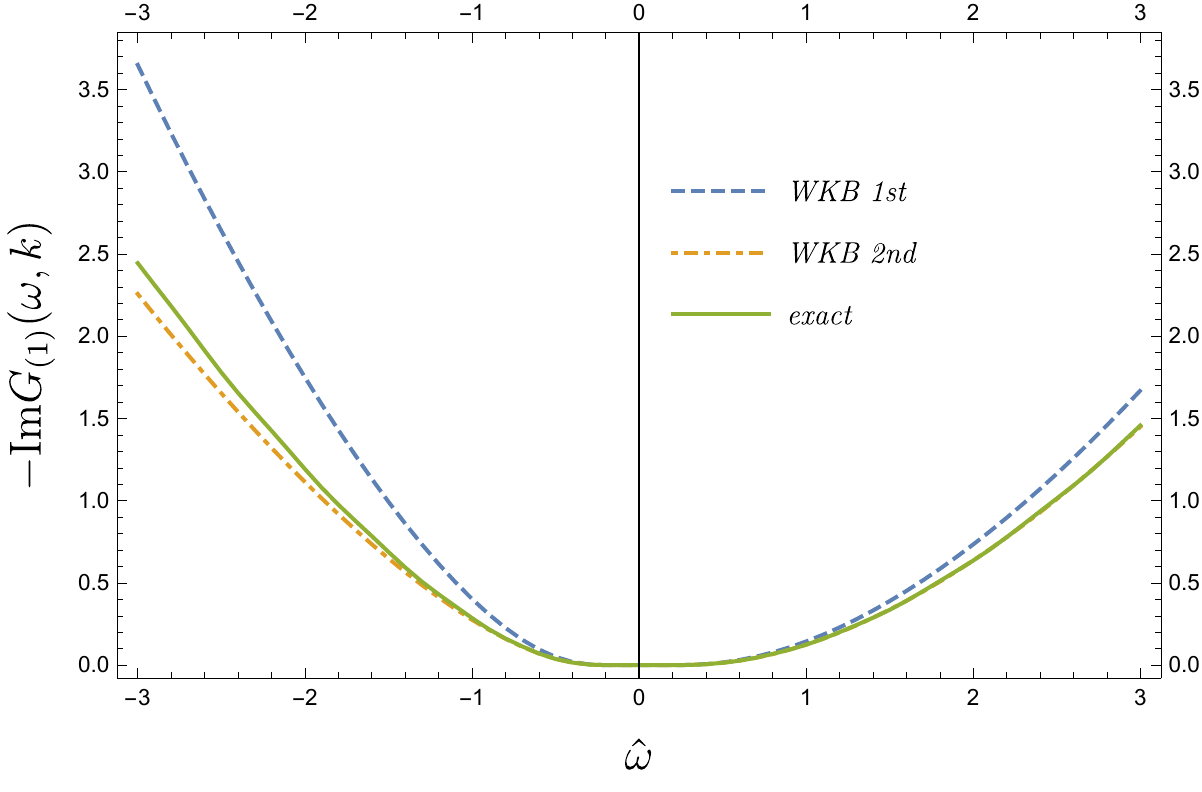}
\caption{Plot of $-\text{Im}G_{(1)}(\omega,k)$ with $z=2$, $m=1.55$, and $k=1$. The three lines represent calculations from first-order WKB, second-order WKB, and the exact numerical solution. The flat region near $\hat{\omega} \approx 0$ features the exponential suppression. One may obtain a plot of $-\text{Im}G_{(2)}(\omega,k)$ by flipping $\hat{\omega}\rightarrow-\hat{\omega}.$ }
\label{fig:g1}
\end{figure}
%%%%

\subsection{Exponential suppression in the $\hat\omega\to0$ limit}

Although the WKB potential (\ref{eq:xpot}) is somewhat complicated, as already seen in (\ref{eq:Vscalar}),
the first three terms match that of the bulk scalar.  In fact, in the limit $\hat\omega\to0$, only these first
three terms are important, and the spin-1/2 potential reduces to that of the spin-0 case.  As a result,
it immediately follows that the exponential suppression of spectral weight in the zero frequency limit
is identical for spin-0 and spin-1/2 operators.

In particular, a matched asymptotic expansion gives the result \cite{Faulkner:2010tq}
\begin{equation}
\mbox{Im}\,G_{(1,2)}\approx-\fft{e^{2\nu_\pm}}2\left(\fft{k}{2\nu_\pm}\right)^{2m}
\exp\left(-\fft{\sqrt\pi\Gamma(1/2\delta)}{z\hat\omega^{1/\delta}\Gamma(z/2\delta)}\right),
\end{equation}
in the limit $\hat\omega\ll\nu^{-\delta}$, where we recall that $\delta=z-1$.  The suppression factor agrees with that of the spin-0 case, (\ref{eq:chitun}),
as it must, since the WKB potential is identical in the zero frequency limit.  The asymptotic behavior in
this limit also agrees with the Green's function extracted from the numerical solution of the Dirac equation,
(\ref{eq:1order}) (see Fig.~\ref{fig:g1}).

%%%%%
\section{Discussion}
\label{sec:dis}

While it may be natural to generalize relativistic holography to the non-relativistic case by, for example,
introducing a Lifshitz bulk of the form (\ref{eq:lifmet}), the lack of boost invariance can have a profound
effect on the dynamics of AdS/CFT.  At a classical level, the fact that non-radial null geodesics no longer
reach the boundary suggests a decoupling of the bulk and boundary, leading to `trapped' modes that
cannot be probed by local observables in the boundary field theory.  This classical intuition was confirmed
in \cite{Keeler:2013msa,Keeler:2014lia} for the case of bulk scalars, and, as we have shown above,
generalizes to bulk fermions as well.

In both scalar and fermion cases, at leading order, the bulk fields satisfy an effective Schr\"odinger equation
with a potential of the form
\begin{equation}
\hat V(x)=\fft{\nu^2}{x^2}-\hat\omega^2x^{2(z-1)}+1+\cdots.
\end{equation}
As we approach the boundary ($x\to0$), the first term in the potential dominates, and the
wavefunction is a power-law, as appropriate for scale-invariant behavior.  On the other hand,
the horizon ($x\to\infty$) behavior is governed by the second term.  Provided $z>1$ (i.e.\ the
non-relativistic case), the potential drops to $-\infty$, and the wavefunction is oscillatory.
For $\hat\omega\ll\nu^{-(z-1)}$, an intermediate region develops with $\hat V\approx1$.  Since
the effective potential is referenced to zero energy, this gives rise to a tunneling barrier, and
elementary considerations demonstrates that the tunneling amplitude is of the form $e^{-s}$
where $s\approx-2/\hat\omega^{1/(z-1)}$.  A more careful WKB analysis the gives rise to the
expression (\ref{eq:chitun}).

Based on the experience with the effective Schr\"odinger potential for spin-0 and spin-1/2 fields
in the bulk, we expect that the result is universal and independent of spin.  This can be verified
for the case of arbitrary integer spin.  Consider, for example, a spin-$\ell$ field described by a
completely symmetric and trace-free field $\Phi_{(M_1M_2\cdots M_\ell)}$ satisfying
\begin{equation}
(\Box-m^2)\Phi_{M_1M_2\cdots M_\ell}=0,\qquad\nabla^N\Phi_{NM_2\cdots M_\ell}=0,\qquad
\Phi^N{}_{NM_3\cdots M_\ell}=0.
\end{equation}
Specializing to the Lifshitz metric (\ref{eq:lifmet}) and taking only the space components
$\Phi_{i_1\cdots i_\ell}$ non-vanishing gives the equation of motion
\begin{equation}
\left[\partial_r^2-\fft{n+z-2\ell-1}r\partial_r-\fft{m^2+\ell(n+z-\ell+1)}{r^2}+\omega^2r^{2(z-1)}-k^2\right]
\Phi_{i_1\cdots i_\ell}=0,
\end{equation}
along with the transverse condition $k^j\Phi_{ji_2\cdots i_\ell}=0$.
This can be converted to Schr\"odinger form by defining
\begin{equation}
\Psi_{i_1\cdots i_\ell}=r^{-(n+z-2\ell-1)/2}\Phi_{i_1\cdots i_\ell}.
\end{equation}
The effective Schr\"odinger equation is then $(-\partial_r^2+V_\ell)\Psi_{i_1\cdots i_\ell}=0$ where
\begin{equation}
V_\ell=\fft{\nu_\ell^2-1/4}{r^2}-\omega^2 r^{2(z-1)}+k^2,\qquad\nu_\ell^2=\ell+\left(\fft{n+z}2\right)^2+m^2.
\label{eq:Schl}
\end{equation}
This reduces to the scalar case (\ref{eq:Vscalar}) when $\ell=0$.

Although the scaling dimension $\nu_\ell$ (or equivalently bulk mass) in (\ref{eq:Schl}) governs the
power-law behavior of the Green's function, the exponential suppression arises from the interplay
between the $\omega^2$ and $k^2$ terms, and hence
is independent of $\nu_\ell$.  In particular, the low-frequency suppression
of spectral weight depends only on the critical exponent $z$ and the scale-invariant quantity $\hat\omega$.
This matches the expectation that the exponential suppression is universal feature of Lifshitz scaling
and is independent of the detailed dynamics of the field theory operators or their bulk dual fields.

Of course, the effective potential (\ref{eq:xpot}), which arises from squaring the Dirac equation, includes
additional terms that are not present for integer spin fields.  Note, however, that they drop out in the relativistic
($z=1$) case, so the potential becomes independent of spin.  This is easily understood from relativistic
conformal invariance, which uniquely determines the Green's function, at least up to kinematical factors
related to spin.  Moving away from $z=1$, on the other hand, relaxes the form of the Green's function,
and in particular allows it to have the behavior $G\sim k^{2\nu z} f(\hat\omega)$, where
$f(\hat\omega)$ is a function of the scale-invariant quantity $\hat\omega$.  Although the
solution to the Dirac equation in a pure Lifshitz background fully determines this function, it will
receive modifications once we introduce higher derivative corrections in the bulk \cite{Keeler:2015afa}.

Just as in the scalar case, such correction terms will relax the form of the Green's function.  Nevertheless,
the exponential suppression of spectral weight at low frequencies is universal, and remains robust, even
in the presence of higher derivative corrections, so long as $\hat\omega\gg(\ell_*/L)^{z-1}$.  Here
$L$ is the Lifshitz curvature scale in (\ref{eq:lifmet}) and $\ell_*$ sets the scale of the higher derivative
expansion.

Finally, although we have focused on the scale-invariant case, our analysis carries over to more general
bulk geometries with broken boost invariance.  The main features of the holographic Green's function can
be seen by the shape of the effective Schr\"odinger potential (\ref{eq:Spotd}), and in particular the
presence or absence of any tunneling barriers.  Unlike for the scalar case, where the dependence of the
effective potential on $\omega$ and $k$ is straightforward, the Dirac case has a rather complicated
behavior in these parameters.  For Lifshitz backgrounds, the additional terms in (\ref{eq:Spotd}) are never
dominant.  However, this is not always the case for more general backgrounds.  As a result, the scalar and
fermion Green's functions can exhibit rather different behavior; this allows for such features as holographic
fermi surfaces that are not present in the bosonic case.  It would be interesting to explore and classify the
structure of potential barriers in the more general setting and to relate the resulting suppression of spectral
weight to the physics of the dual field theory.

%%%%%%%%%%
\section*{Acknowledgments}

JTL wishes to thank C.\ Keeler, G.\ Knodel and K.\ Sun for useful discussions.
This work was supported in part by the US Department of Energy under grant DE-SC0007859.

%%%%%%%%%%

\appendix

\section{The WKB approximation for $1/r^2$ potentials}
\label{app:WKB}

In this appendix, we review the second order WKB approximation and furthermore apply it to
the case of a $1/r^2$ potential.  Our starting point is the second order equation
\begin{equation}
-\epsilon^2y''(x)+Q(x)y(x)=0,
\label{eq:sch}
\end{equation}
where we consider $\epsilon$ to be our expansion parameter.  We then expand
\begin{equation}
y(x)\sim\exp\left[\fft1\epsilon\sum_{n\ge0}\epsilon^nS_n(x)\right].
\end{equation}
Substituting the series expansion into (\ref{eq:sch}) and matching powers of $\epsilon$ gives
\begin{align}
S_0'&=\pm\sqrt{Q},\nn\\
S_1'&=-\fft{S_0''}{2S_0'}=-\fft12\fft{d}{dx}\log(S_0'),\nn\\
S_2'&=-\fft{S_1''+S_1'^2}{2S_0'},
\label{eq:Sexp}
\end{align}
and so on.  In general, the terms $S_n$ are total derivatives for odd $n$, but not for even $n$.  Integrating
$S_0'$ and $S_2'$ allows us to write the second order WKB wavefunction
\begin{equation}
y(x)\sim Q^{-1/4}\exp\left[\pm\fft1\epsilon\int^x\sqrt{Q}dx\pm\epsilon\left(\fft{5Q'}{48Q^{3/2}}
+\int^x\fft{Q''}{48Q^{3/2}}dx\right)+\mathcal O(\epsilon^2)\right],
\end{equation}
where we have integrated $S_2'$ by parts.  Ignoring the $\mathcal O(\epsilon)$ term in the
exponent gives the usual first order WKB approximation.

\subsection{Connection formulae}
The above expression is general for both classically forbidden ($Q>0$) and classically allowed ($Q<0$)
regions, provided we allow the factors to become complex.  (Up to second order, the exponent is a pure
phase in the classically allowed region, although this does persist at higher orders.)  Since the approximation
breaks down at a turning point, we use a matched asymptotic expansion in order to connect solutions
across the turning point.   For the effective Schr\"odinger potentials we are interested in,
we take $Q$ positive for $x<0$ and $Q$ negative for $x>0$.  Here
we have chosen the turning point to be $x=0$ for simplicity.  

To fix the constants, we take
\begin{equation}
y(x)=Q^{-1/4}\left\{\begin{matrix}F\\G\end{matrix}\right\}\exp\left[\pm\fft1\epsilon\int_x^0\sqrt{Q}dx
\mp\epsilon\left(\fft{5Q'}{48Q^{3/2}}-\int_x^{-\mu}\fft{Q''}{48Q^{3/2}}dx
+\fft{\beta}{12(-\alpha)^{3/2}\sqrt\mu}\right)\right],
\label{eq:y1}
\end{equation}
on the left (i.e.\ for $x<0$).  The last factor requires some explanation.  Since the integrand $Q''/Q^{3/2}$
is singular at the turning point, we cannot integrate all the way to $x=0$.  Instead, we regulate the integral
by a small cutoff $\mu\to0^+$.  The last term then cancels the divergence where $a$ and $b$ are Taylor
coefficients near the turning point
\begin{equation}
Q(x)=\alpha x+\beta x^2+\cdots\qquad\mbox{for }x\to0.
\end{equation}
Note that $\alpha<0$ since $Q$ has a negative slope going through the turning point.

For the solution on the right ($x>0$), we take
\begin{equation}
y(x)=|Q|^{-1/4}\left\{\begin{matrix}a\\b\end{matrix}\right\}\exp\left[\pm\fft{i}\epsilon\int_0^x\!\!\!\sqrt{|Q|}dx
\mp i\epsilon\left(\fft{5|Q|'}{48|Q|^{3/2}}+\int_\mu^x\!\!\fft{|Q|''}{48|Q|^{3/2}}dx
+\fft{\beta}{12(-\alpha)^{3/2}\sqrt\mu}\right)\right].
\label{eq:y3}
\end{equation}
We match the left and right expressions by solving the equation
\begin{equation}
-\epsilon^2y''+(\alpha x+\beta x^2+\cdots)y=0
\end{equation}
near the turning point.  Ignoring $\beta$ gives an Airy function, while including $\beta$ and working to the next order
gives
\begin{equation}
y=\left(1-\fft{\beta x}{5\alpha}+\cdots\right)\left\{\begin{matrix}\mbox{Ai}\\\mbox{Bi}\end{matrix}\right\}
\left(-\epsilon^{-2/3}(-\alpha)^{1/3}x\left(1+\fft{\beta x}{5\alpha}+\cdots\right)\right).
\label{eq:y2}
\end{equation}
This expression is for $\alpha<0$.  (For $\alpha>0$, the initial part of the argument of the Airy function should be
replaced by $+\epsilon^{-2/3}\alpha^{1/3}x$.)

The connection formula between the forbidden and allowed solutions may be obtained by asymptotic
expansion of (\ref{eq:y1}), (\ref{eq:y3}) and (\ref{eq:y2}).  It turns out that the expression is
identical to that of first order WKB
\begin{equation}
\begin{pmatrix}F\\G\end{pmatrix}=\begin{pmatrix}e^{-i\pi/4}&e^{i\pi/4}\\\fft12e^{i\pi/4}&\fft12e^{-i\pi/4}\end{pmatrix}
\begin{pmatrix}a\\b\end{pmatrix}.
\end{equation}

\subsection{Handling the singular $1/r^2$ potential}

In order for WKB to be a controlled expansion, we want the terms $S_n$ to be well behaved as an
asymptotic series.  This fails when we have $Q(x)\sim1/x^2$.  Consider, for example $Q=(\tilde\nu/x)^2$.
Then the expansion (\ref{eq:Sexp}) gives
\begin{equation}
S_0=\tilde\nu\log x,\qquad S_1=\fft12\log x,\qquad S_2=\fft1{8\tilde\nu}\log x,
\end{equation}
where we have taken the positive root.  In this case, the higher order terms are just as important as
the first order terms, and cannot systematically be dropped.

In order to handle a $1/r^2$ potential, consider the Schr\"odinger equation
\begin{equation}
-\fft{d^2\psi(r)}{dr^2}+V(r)\psi(r)=0.
\label{eq:schr}
\end{equation}
We now make the transformation
\begin{equation}
r=e^x,\qquad \psi(r)=e^{x/2}y(x).
\end{equation}
This transforms the Schr\"odinger equation into
\begin{equation}
-y''(x)+Q(x)y(x)=0
\end{equation}
where
\begin{equation}
Q(x)=\left.r^2V(r)+\fft14\right|_{r=e^x}
\label{eq:QrV}
\end{equation}
We have thus transformed the singular potential into an asymptotically constant one
\begin{equation}
V(r)\sim\fft{\nu^2-\fft14}{r^2}\mbox{ as }r\to0\qquad\Rightarrow\qquad Q(x)\sim\nu^2\mbox{ as }x\to-\infty.
\end{equation}
This allows us to use the WKB approximation for the $Q(x)$ potential.

Transforming $y(x)$ in (\ref{eq:y1}) back to $\psi(r)$ gives
\begin{align}
\psi(r)=\hat V^{-1/4}\left\{\begin{matrix}F\\G\end{matrix}\right\}\exp\biggl[
&\pm\fft1\epsilon\int_r^{r_0}\sqrt{\hat V}dr\nn\\
&\mp\epsilon\biggl(\fft{5(\hat V'+\fft2r\hat V)}{48\hat V^{3/2}}
-\int_r^{r_0-\mu}\fft{\hat V''+\fft5r\hat V'+\fft4{r^2}\hat V}{48\hat V^{3/2}}dr
+\fft{\fft{5\alpha}{2r_0}+\beta}{12(-\alpha)^{3/2}\mu^{1/2}}\biggr)\biggr],
\end{align}
where
\begin{equation}
\hat V=V+\fft1{4r^2}.
\end{equation}
Here $r_0$ is the classical turning point and
\begin{equation}
\hat V=\alpha(r-r_0)+\beta(r-r_0)^2+\cdots\quad\mbox{as}\quad r\to r_0.
\end{equation}
In particular, $\alpha=\hat V'(r_0)$ and $\beta=\hat V''(r_0)/2$.

One thing to note is that the
terms in the small $\epsilon$ expansion are rearranged between the $\psi(r)$ equation and the
$y(x)$ equation.  If we put in an explicit $\epsilon^2$ in the second derivative term of the
Schr\"odinger equation (\ref{eq:schr}), then the $1/4$ term in the potential $Q(x)$ of (\ref{eq:QrV})
would become $\epsilon^2/4$.  However, since we are interested in $\epsilon=1$, the $\epsilon$
independent shift by $1/4$ is appropriate.

%%%%%%%%%%
\section{The Green's function from the numerical method}
\label{app:exact}

In this appendix, we explain the procedure to compute the imaginary part of the Green's function from the numerical solution of the Dirac equation.  We follow the approach of 
\cite{Iqbal:2008by,Liu:2009dm} and convert the coupled linear equations resulting from the components of
the Dirac equation into a first order Riccati flow equation.

We defined $\eta$ and $\xi$ in (\ref{eq:fetaxi}) and showed that their projections, $\eta_\pm = P_\pm \eta$ and $\xi_\pm = P_\pm \xi$, follow the pair of coupled equations (\ref{eq:1order}), which we repeat here:
\begin{eqnarray}
(\partial_r+m/r)\eta_\pm&=&(\omega r^{z-1}\mp k)\xi_\pm,\nn\\
(\partial_r-m/r)\xi_\pm&=&-(\omega r^{z-1}\pm k)\eta_\pm.
\label{eq:1orderrep}
\end{eqnarray}
The boundary behavior of $\eta_\pm$ and $\xi_\pm$ is
\begin{eqnarray}
\eta_\pm &=& \alpha_\pm r^{-m} + \beta_\pm r^{m+1},\nn\\
\xi_\pm &=& \gamma_\pm r^{-m+1} + \delta_\pm r^{m},
\end{eqnarray}
as $r \rightarrow 0$, while the horizon behavior is
\begin{equation}
\eta_\pm = ae^{\pm i\omega r^z/z}, \quad\xi_\pm = \pm iae^{\pm i\omega r^z/z},
\end{equation}
as $r \rightarrow \infty$.

Although $\eta_\pm$ and $\xi_\pm$ are spinors, their equations of motion (\ref{eq:1orderrep}) do not contain matrices, i.e., do not mix the spinor components. Therefore, we may consider each of these spinors as a single function. This allows us to define the ratio $\zeta_\pm = \xi_\pm / \eta_\pm$. The differential equation governing this ratio $\zeta_\pm$ is given by
\beq
\partial_r \zeta_\pm + (\omega r^{z-1} \mp k ) \zeta_\pm^2 - \frac{2m}{r}\zeta_\pm + (\omega r^{z-1} \pm k) = 0.
\label{eq:zetaeq}
\eeq
This ratio $\zeta_\pm$ exhibits boundary asymptotic behavior
\beq
\zeta_\pm = \frac{\gamma_\pm}{\alpha_\pm} r + \frac{\delta_\pm}{\alpha_\pm} r^{2m},
\label{eq:zetabd}
\eeq
as $r \rightarrow 0$, and horizon asymptotic behavior
\beq
\zeta_\pm = i,\quad\mbox{or}\quad \zeta_\pm = -i
\eeq
as $r \rightarrow \infty$. We will take $\zeta_\pm$ to be $i$ at the horizon, which corresponds to the infalling boundary condition. We note that each pair $\{ \alpha_\pm, \gamma_\pm \}$ and $\{ \beta_\pm, \delta_\pm \}$ in (\ref{eq:zetabd}) is not independent. They are related by the coupled equations (\ref{eq:1orderrep}) as
\begin{eqnarray}
\gamma_\pm=\frac{\mp k}{- 2m+1}\alpha_\pm, \quad \beta_\pm=\frac{\mp k}{2m+1}\delta_\pm
\end{eqnarray}
in the limit $r \rightarrow 0$. The first relation shows that the ratio $\gamma_\pm/\alpha_\pm$ is purely real. This means that we can pull out the imaginary part of the $\delta_\pm/\alpha_\pm$ term by
\beq
\mbox{Im} \frac{\delta_\pm}{\alpha_\pm} = \lim_{r\rightarrow0} \mbox{Im} \,\zeta_\pm r^{-2m}.
\label{eq:zetalim}
\eeq
This quantity can be used to obtain the imaginary part of the Green's function as done in section \ref{sec:extrac}. Using the definition (\ref{eq:fetaxi}) to obtain $f_\pm$ and comparing to the asymptotic equations (\ref{eq:asybeh}), we find
\beq
A = \ft{1}{2}(1-\Gamma^{\bar{r}})(P_+\alpha_- + P_-\alpha_+), \quad D = -i\Gamma^{\bar{t}} \ft{1}{2}(1-\Gamma^{\bar{r}})(P_+ \delta_- + P_- \delta_+),
\eeq
where we show the projection explicitly. The Green's function relates these quantities as specified by the equation (\ref{eq:Gdef}), from which we obtain the imaginary part of the Green's function
\beq
\mbox{Im} G_R = -\mbox{Im}\, \fft{1}{2}(1+\Gamma^{\bar{r}})(P_+ \frac{\delta_-}{\alpha_-} + P_- \frac{\delta_+}{\alpha_+}).
\label{eq:imgreen}
\eeq
In brief, we have the following procedure to obtain the imaginary part of the Green's function. We seek the numerical solution of the differential equation for $\zeta_\pm$, (\ref{eq:zetaeq}), with the initial condition, $\zeta_\pm(r\rightarrow\infty) = i$. Then we evaluate $\zeta_\pm(r\rightarrow0)$ to obtain the ratio $\mbox{Im} \,\delta_\pm/\alpha_\pm$ from (\ref{eq:zetalim}). Finally, the expression (\ref{eq:imgreen}) gives us $\mbox{Im}\, G_R$.

%%%%%%%%

%%%%%%%%%%

%%%%%%%%%%

\begin{thebibliography}{99}

\bibitem{Kachru:2008yh}
S.~Kachru, X.~Liu and M.~Mulligan,
{\sl Gravity duals of Lifshitz-like fixed points},
Phys.\ Rev.\ D {\bf 78}, 106005 (2008) [arXiv:0808.1725 [hep-th]].
%%CITATION = doi:10.1103/PhysRevD.78.106005;%%

\bibitem{Hartnoll:2009sz}
S.~A.~Hartnoll,
{\sl Lectures on Holographic Methods for Condensed Matter Physics},
Class.\ Quant.\ Grav.\ {\bf 26} (2009) 224002 [arXiv:0903.3246 [hep-th]].
%%CITATION = doi:10.1088/0264-9381/26/22/224002;%%

\bibitem{Copsey:2010ya}
K.~Copsey and R.~Mann,
{\sl Pathologies in Asymptotically Lifshitz Spacetimes},
JHEP {\bf 1103} (2011) 039 [arXiv:1011.3502 [hep-th]].
%%CITATION = doi:10.1007/JHEP03(2011)039;%%

\bibitem{Horowitz:2011gh}
G.~T.~Horowitz and B.~Way,
{\sl Lifshitz Singularities},
Phys.\ Rev.\ D {\bf 85}, 046008 (2012) [arXiv:1111.1243 [hep-th]].
%%CITATION = ARXIV:1111.1243;%%

\bibitem{Harrison:2012vy}
S.~Harrison, S.~Kachru and H.~Wang,
{\sl Resolving Lifshitz Horizons},
JHEP {\bf 1402}, 085 (2014) [arXiv:1202.6635 [hep-th]].
%%CITATION = ARXIV:1202.6635;%%

\bibitem{Bao:2012yt}
N.~Bao, X.~Dong, S.~Harrison and E.~Silverstein,
{\sl The Benefits of Stress: Resolution of the Lifshitz Singularity},
Phys.\ Rev.\ D {\bf 86} (2012) 106008 [arXiv:1207.0171 [hep-th]].
%%CITATION = doi:10.1103/PhysRevD.86.106008;%%

\bibitem{Bhattacharya:2012zu}
J.~Bhattacharya, S.~Cremonini and A.~Sinkovics,
{\sl On the IR Completion of Geometries with Hyperscaling Violation},
JHEP {\bf 1302} (2013) 147 [arXiv:1208.1752 [hep-th]].
%%CITATION = doi:10.1007/JHEP02(2013)147;%%

\bibitem{Knodel:2013fua}
G.~Knodel and J.~T.~Liu,
{\sl Higher derivative corrections to Lifshitz backgrounds},
JHEP {\bf 1310}, 002 (2013) [arXiv:1305.3279 [hep-th]].
%%CITATION = ARXIV:1305.3279;%%

\bibitem{Keeler:2013msa}
C.~Keeler, G.~Knodel and J.~T.~Liu,
{\sl What do non-relativistic CFTs tell us about Lifshitz spacetimes?},
JHEP {\bf 1401}, 062 (2014) [arXiv:1308.5689 [hep-th]].
%%CITATION = ARXIV:1308.5689;%%

\bibitem{Faulkner:2010tq}
T.~Faulkner and J.~Polchinski,
{\sl Semi-Holographic Fermi Liquids},
JHEP {\bf 1106}, 012 (2011) [arXiv:1001.5049 [hep-th]].
%%CITATION = doi:10.1007/JHEP06(2011)012;%%

\bibitem{Hartnoll:2012wm}
S.~A.~Hartnoll and E.~Shaghoulian,
{\sl Spectral weight in holographic scaling geometries},
JHEP {\bf 1207}, 078 (2012) [arXiv:1203.4236 [hep-th]].
%%CITATION = doi:10.1007/JHEP07(2012)078;%%

\bibitem{Keeler:2014lia}
C.~Keeler, G.~Knodel and J.~T.~Liu,
{\sl Hidden horizons in non-relativistic AdS/CFT},
JHEP {\bf 1408}, 024 (2014) [arXiv:1404.4877 [hep-th]].
%%CITATION = ARXIV:1404.4877;%%

\bibitem{Hartnoll:2012rj}
S.~A.~Hartnoll and D.~M.~Hofman,
{\sl Locally Critical Resistivities from Umklapp Scattering},
Phys.\ Rev.\ Lett.\  {\bf 108}, 241601 (2012) [arXiv:1201.3917 [hep-th]].

\bibitem{Henningson:1998cd}
M.~Henningson and K.~Sfetsos,
{\sl Spinors and the AdS / CFT correspondence},
Phys.\ Lett.\ B {\bf 431}, 63 (1998) [hep-th/9803251].

\bibitem{Mueck:1998iz}
W.~Mueck and K.~S.~Viswanathan,
{\sl Conformal field theory correlators from classical field theory on anti-de Sitter space. 2. Vector and spinor fields},
Phys.\ Rev.\ D {\bf 58}, 106006 (1998) [hep-th/9805145].

\bibitem{Henneaux:1998ch}
M.~Henneaux,
{\sl Boundary terms in the AdS / CFT correspondence for spinor fields},
hep-th/9902137.

\bibitem{Iqbal:2009fd}
N.~Iqbal and H.~Liu,
{\sl Real-time response in AdS/CFT with application to spinors},
Fortsch.\ Phys.\  {\bf 57}, 367 (2009) [arXiv:0903.2596 [hep-th]].
  %%CITATION = ARXIV:0903.2596;%%

\bibitem{Liu:2009dm}
H.~Liu, J.~McGreevy and D.~Vegh,
{\sl Non-Fermi liquids from holography},
Phys.\ Rev.\ D {\bf 83}, 065029 (2011) [arXiv:0903.2477 [hep-th]].

\bibitem{Faulkner:2009wj}
T.~Faulkner, H.~Liu, J.~McGreevy and D.~Vegh,
{\sl Emergent quantum criticality, Fermi surfaces, and AdS$_2$},
Phys.\ Rev.\ D {\bf 83}, 125002 (2011) [arXiv:0907.2694 [hep-th]].

\bibitem{Iizuka:2011hg}
N.~Iizuka, N.~Kundu, P.~Narayan and S.~P.~Trivedi,
{\sl Holographic Fermi and Non-Fermi Liquids with Transitions in Dilaton Gravity},
JHEP {\bf 1201}, 094 (2012) [arXiv:1105.1162 [hep-th]].
%%CITATION = doi:10.1007/JHEP01(2012)094;%%

\bibitem{Hartnoll:2011dm}
S.~A.~Hartnoll, D.~M.~Hofman and D.~Vegh,
{\sl Stellar spectroscopy: Fermions and holographic Lifshitz criticality},
JHEP {\bf 1108}, 096 (2011) [arXiv:1105.3197 [hep-th]].
%%CITATION = doi:10.1007/JHEP08(2011)096;%%

\bibitem{Iqbal:2011in}
N.~Iqbal, H.~Liu and M.~Mezei,
{\sl Semi-local quantum liquids},
JHEP {\bf 1204}, 086 (2012) [arXiv:1105.4621 [hep-th]].
%%CITATION = doi:10.1007/JHEP04(2012)086;%%

\bibitem{Cubrovic:2011xm}
M.~Cubrovic, Y.~Liu, K.~Schalm, Y.~W.~Sun and J.~Zaanen,
{\sl Spectral probes of the holographic Fermi groundstate: dialing between the electron star and AdS Dirac hair},
Phys.\ Rev.\ D {\bf 84}, 086002 (2011) [arXiv:1106.1798 [hep-th]].
%%CITATION = doi:10.1103/PhysRevD.84.086002;%%

\bibitem{Gursoy:2012ie}
U.~Gursoy, V.~Jacobs, E.~Plauschinn, H.~Stoof and S.~Vandoren,
{\sl Holographic models for undoped Weyl semimetals},
JHEP {\bf 1304}, 127 (2013) [arXiv:1209.2593 [hep-th]].
%%CITATION = doi:10.1007/JHEP04(2013)127;%%

\bibitem{Alishahiha:2012nm}
M.~Alishahiha, M.~R.~Mohammadi Mozaffar and A.~Mollabashi,
{\sl Fermions on Lifshitz Background},
Phys.\ Rev.\ D {\bf 86}, 026002 (2012) [arXiv:1201.1764 [hep-th]].
%%CITATION = doi:10.1103/PhysRevD.86.026002;%%

\bibitem{Fang:2012pw}
L.~Q.~Fang, X.~H.~Ge and X.~M.~Kuang,
{\sl Holographic fermions in charged Lifshitz theory},
Phys.\ Rev.\ D {\bf 86}, 105037 (2012) [arXiv:1201.3832 [hep-th]].
%%CITATION = doi:10.1103/PhysRevD.86.105037;%%

\bibitem{Son:2002sd}
D.~T.~Son and A.~O.~Starinets,
{\sl Minkowski space correlators in AdS / CFT correspondence: Recipe and applications},
JHEP {\bf 0209}, 042 (2002) [hep-th/0205051].
%%CITATION = doi:10.1088/1126-6708/2002/09/042;%%

\bibitem{Horowitz:2009ij}
G.~T.~Horowitz and M.~M.~Roberts,
{\sl Zero Temperature Limit of Holographic Superconductors},
JHEP {\bf 0911}, 015 (2009) [arXiv:0908.3677 [hep-th]].
%%CITATION = doi:10.1088/1126-6708/2009/11/015;%%

\bibitem{Basu:2009xf}
P.~Basu,
{\sl Energy scales in a holographic black hole and conductivity at finite momentum},
Can.\ J.\ Phys.\  {\bf 89}, 271 (2011) [arXiv:0911.5082 [hep-th]].
%%CITATION = doi:10.1139/P11-019;%%

\bibitem{Keeler:2015afa}
C.~Keeler, G.~Knodel, J.~T.~Liu and K.~Sun,
{\sl Universal features of Lifshitz Green's functions from holography},
JHEP {\bf 1508}, 057 (2015) [arXiv:1505.07830 [hep-th]].
%%CITATION = ARXIV:1505.07830;%%

\bibitem{Iqbal:2008by}
N.~Iqbal and H.~Liu,
{\sl Universality of the hydrodynamic limit in AdS/CFT and the membrane paradigm},
Phys.\ Rev.\ D {\bf 79}, 025023 (2009) [arXiv:0809.3808 [hep-th]].
%%CITATION = doi:10.1103/PhysRevD.79.025023;%%


\end{thebibliography}
\end{document}